\documentclass[aps,prb,twocolumn]{revtex4}
\usepackage{graphicx}
\usepackage{natbib}

\begin{document}

\title{
Conformational and Structural Relaxations of Poly(ethylene oxide) and Poly(propylene
oxide) Melts: Molecular Dynamics Study of Spatial Heterogeneity, Cooperativity, and
Correlated Forward-Backward Motion}

\author{Michael Vogel}
\affiliation{Institut f\"ur Physikalische Chemie, Westf\"alische Wilhelms-Universit\"at
M\"unster, Corrensstr.\ 30/36, 48149 M\"unster, Germany}

\date{\today}

\begin{abstract}

Performing molecular dynamics simulations for all-atom models, we characterize the
conformational and structural relaxations of poly(ethylene oxide) and poly(propylene
oxide) melts. The temperature dependence of these relaxation processes deviates from an
Arrhenius law for both polymers. We demonstrate that mode-coupling theory captures some
aspects of the glassy slowdown, but it does not enable a complete explanation of the
dynamical behavior. When the temperature is decreased, spatially heterogeneous and
cooperative translational dynamics are found to become more important for the structural
relaxation. Moreover, the transitions between the conformational states cease to obey
Poisson statistics. In particular, we show that, at sufficiently low temperatures,
correlated forward-backward motion is an important aspect of the conformational
relaxation, leading to strongly nonexponential distributions for the waiting times of the
dihedrals in the various conformational states.

\end{abstract}
\maketitle

\section{Introduction}

Understanding the glass transition of polymer melts is of enormous interest from the
viewpoints of fundamental and applied science. When polymer melts or other glass-forming
liquids are cooled, the structural relaxation slows down tremendously and, eventually, an
amorphous solid, a glass, is formed at the glass transition temperature $T_g$. The
structural relaxation, or, equivalently, the $\alpha$ relaxation of most glass-forming
liquids exhibits two striking features: its time dependence differs from a single
exponential function and its temperature dependence deviates from an Arrhenius
law.\cite{Ed_96}

Molecular dynamics (MD) simulations have proven a powerful tool to investigate the
initial stages of the glassy slowdown of the structural relaxation in moderately viscous
liquids.\cite{Bi_00,Gl_00_1,De_01,Pa_04,Ba_05} Most of these studies focused on simple
model-glass formers, e.g., binary Lennard-Jones mixtures. In MD simulations of
glass-forming polymer melts, coarse-grained models were used, while investigations on
chemically realistic models are more rare.\cite{Pa_04,Ba_05} For the latter models,
consideration of all the interactions associated with the connectivity, e.g., of the
energy barriers against the conformational dynamics, interferes with the necessity to
follow the slowdown of the molecular dynamics over broad temperature and time ranges due
to the limited computer power. However, the rapid development of computer technology
steadily improves the possibilities and, very recently, chemically realistic polymer
models have started to become an important tool to study the mechanisms for the primary
and Johari-Goldstein secondary relaxation processes in polymer melts.\cite{Sm_07,Co_07}

The mode-coupling theory (MCT), which focuses on density fluctuations, has been put
forward to explain the observation that the temperature dependent structural relaxation
times deviate from an Arrhenius law for most glass-forming liquids.\cite{Go_92} In its
idealized version, MCT predicts a power-law divergence of the $\alpha$-relaxation time at
a critical temperature $T_c$. MD simulations were employed to test the predictions of MCT
for moderately viscous liquids. It was found that this theory captures many aspects of
the glassy slowdown for several simple model-glass formers, including a bead-spring
polymer model,\cite{Ba_05,Ko_95,Ho_01} while the applicability to all-atom polymer models
was controversially discussed.\cite{Kr_03,Pa_04,Co_07,Pa_06} In any event, $T_c$ is
substantially higher than $T_g$, indicating that MCT fails to describe the molecular
dynamics in highly viscous liquids.\cite{Ed_96} Therefore, it is important to consider
further aspects of glass-forming liquids and approaches focusing on the cooperativity and
heterogeneity of the structural relaxation have received considerable
attention.\cite{Ad_65,Ga_02}

Various experimental observations demonstrated that the nonexponential $\alpha$
relaxation of glass-forming liquids is related to the effect that the molecular dynamics
are heterogeneous, i.e., it is possible to select particles that rotate or translate much
farther or shorter distances than an average particle.\cite{Bo_98,Si_99,Ed_00} However,
most experimental techniques provide only limited information about the spatial
distribution of particles showing different mobilities. Recent nuclear magnetic resonance
approaches demonstrated for various glass-forming liquids, including a polymer melt, that
the dynamics are spatially heterogeneous.\cite{Tr_98,Re_01,Qi_03} Specifically, particles
within a physical region of the liquid show an enhanced or diminished mobility as
compared to particles in a region a few nanometers away. Nevertheless, a detailed
experimental characterization of the time and temperature dependence of spatially
heterogenous dynamics is still lacking.

MD simulations provide straightforward access to spatial correlations of the particle
mobility.\cite{Gl_00_1,Ba_05} Work on simple models of atomic and polymeric glass-forming
liquids reported that highly mobile and highly immobile particles aggregate into
clusters, which are transient in
nature.\cite{Ko_97,Dol_98,Don_99_1,Don_99_2,Be_99,Gl_00_2,Ge_01,Ge_04} The clusters of
highly mobile particles are largest in the very early stages of the $\alpha$ relaxation.
Upon cooling, the cluster size increases and a divergence near the critical temperature
$T_c$ was proposed.\cite{Don_99_1,Don_99_2,Be_99} Moreover, it was demonstrated that
string-like motion is an important channel for the structural relaxation of the highly
mobile particles.\cite{Don_98,Ai_03,Ge_04} This means that mobile particles tend to
follow each other along one-dimensional paths. The existence of spatially heterogeneous
dynamics is not restricted to simple models, but this effect was also observed for more
complex models of glass forming liquids, namely, for models of propylene carbonate,
water, and silica.\cite{Qi_99,Gi_03,Vo_04_1,Vo_04_2,Te_04} Interestingly, the cooperative
string-like motion was found to be of little relevance for the silicon atoms in the
silica model, where the structural relaxation follows an Arrhenius
law.\cite{Vo_04_1,Vo_04_2}

It was argued that the structural relaxation differs between bead-spring and all-atom
polymer models.\cite{Pa_04} Specifically, conformational dynamics are of central
importance for the latter, but not for the former models. For all-atom polymer models,
the slowdown of the $\alpha$ relaxation was argued to depend not only on intermolecular
packing effects, as assumed in MCT, but also on intramolecular torsional
barriers.\cite{Kr_03,Pa_04,Pa_06,Co_07} Moreover, previous work showed that the
conformational relaxation is by no means a simple relaxation process, but it involves,
e.g., pronounced dynamical heterogeneities.\cite{Sm_07,Bo_00,Bo_03_0} In view of all
these results, it becomes apparent that improving our understanding of the polymer glass
transition requires a detailed characterization of the interplay of the intra- and
intermolecular aspects of the molecular dynamics.

Here, we perform MD simulations for all-atom models of poly(ethylene oxide) (PEO) and
poly(propylene oxide) (PPO). Due to the capability to dissolve salts, PEO and PPO are
popular materials for the preparation of polymer electrolytes.\cite{Gr_91} Therefore, MD
approaches studied how the presence of ions affects the structure and the dynamics of
these polymers both in the bulk and in
confinement.\cite{Bo_00,Ne_95,Mu_95,Ca_95,Li_96,Sm_96,Ah_00,Bo_03_2,Bo_03_3,Ha_00,Ku_03}
Our investigation focuses on the temperature dependent dynamics of the neat model polymer
melts. Specifically, we study the applicability of MCT and the relevance of spatially
heterogeneous and cooperative dynamics for the structural relaxation. Moreover, we
perform a detailed characterization of the conformational relaxation, elucidating the
importance of correlated forward-backward jumps.

\section{Methods}

The studied PEO and PPO models are comprised of polymer chains,
H-[CH$_2$-O-CH$_2$]$_{12}$-H and CH$_3$-O-[CH$_2$-CH(CH$_3$)-O]$_{11}$-CH$_3$,
respectively. Each model is composed of 32 chains. The PPO chains are atactic, i.e., the
side groups are randomly connected. The interatomic interactions of these polymer models
are described by two well established quantum-chemistry based, all-atom force
fields,\cite{Sm_98,Bo_03_1} which can be written in the form
\begin{eqnarray}
V(\{\mathbf{r}\})&=&\sum_{bonds}V^{bo}(r_{ij}) + \sum_{angles}V^{be}(\theta_{ijk}) +
\nonumber\\
&&\sum_{dihedrals}V^{to}(\phi_{ijkl})+V^{nb}(\{\mathbf{r}\})\;.
\end{eqnarray}
Here, $\{\mathbf{r}\}$ is the set of all atomic coordinates. The bonded interactions are
comprised of energies due to stretching of bonds, bending of valence angles, and torsion
of dihedral angles. The nonbonded interactions $V^{nb}$ are composed of Coulombic and van
der Waals interactions, the latter being modeled using a Buckingham potential. The
explicit form of the various interaction terms and the corresponding potential parameters
are given in the literature.\cite{Sm_98,Bo_03_1,FN} In the case of PEO, we apply the
force field termed FF-3 in Ref.\ \onlinecite{Bo_03_1}. It was demonstrated that these
models enable a good reproduction of thermodynamic, structural, and dynamical aspects of
PEO and PPO melts.\cite{Ah_00,Bo_03_2,Bo_03_3}

The MD simulations were performed using the GROMACS software package.\cite{GROM} We
applied periodic boundary conditions and a time step of $1\mathrm{\,fs}$. The nonbonded
interactions were calculated utilizing a cutoff distance of $12\mathrm{\,\AA}$. To treat
the Coulombic interactions, the particle-mesh Ewald technique was employed.\cite{PME} The
LINCS algorithm was used to constrain all bonds.\cite{LINCS} Prior to data acquisition,
the systems were equilibrated in simulations at constant $N$, $P$, and $T$, using the
Rahman-Parrinello barostat\cite{RP} and the Nos\'{e}-Hoover thermostat.\cite{NH} These
equilibration runs, which spanned $20\!-\!30\mathrm{\,ns}$ at the lower temperatures,
allowed us to adjust the densities $\rho(T)$. The density increases from
$\rho(450\mathrm{\,K})\!=\!1.01\mathrm{\,g/cm^3}$ to
$\rho(280\mathrm{\,K})\!=\!1.12\mathrm{\,g/cm^3}$ for PEO and from
$\rho(450\mathrm{\,K})\!=\!0.89\mathrm{\,g/cm^3}$ to
$\rho(250\mathrm{\,K})\!=\!1.05\mathrm{\,g/cm^3}$ for PPO. The subsequent production runs
were performed in the canonical ensemble, i.e., at constant $N$, $V$, and $T$, employing
the Nose-Hoover thermostat. Although experimental work demonstrated that PEO is partially
crystalline at room temperature and ambient pressure,\cite{Gr_91} the simulation results
give no evidence for an onset of crystallization in the studied time and temperature
ranges.

\section{Results}

\subsection{General characterization of the structural
relaxation}\label{sec_gen}

\begin{figure}
\centering
\includegraphics[width=8.25cm]{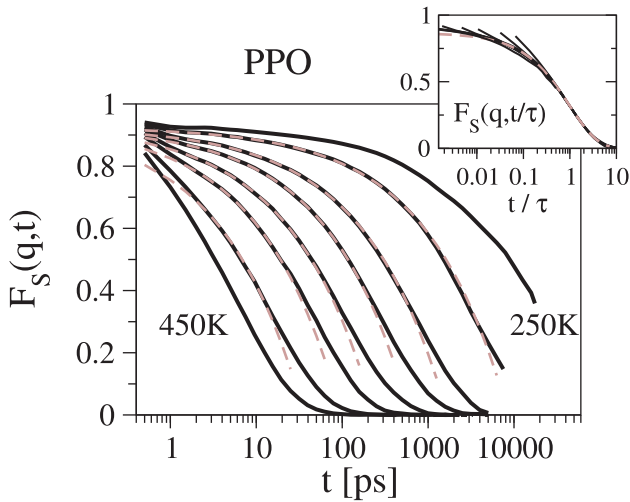}
\caption{\small \textsl{Temperature dependent incoherent intermediate scattering
functions $F_s(q,t)$ for the oxygen atoms of PPO and $q\!=\!0.91\mathrm{\,\AA^{-1}}$. The
temperatures are, from left to right, $450\mathrm{\,K}$, $400\mathrm{\,K}$,
$350\mathrm{\,K}$, $325\mathrm{\,K}$, $300\mathrm{\,K}$, $280\mathrm{\,K}$,
$265\mathrm{\,K}$, and $250\mathrm{\,K}$. The dashed lines are von Schweidler fits, see
Eq.\ (\ref{eq_von}). The inset shows the time temperature superposition of the same data.
The time constants $\tau$ from KWW interpolations of the $\alpha$ relaxation were used
for the scaling of the time axis. The dashed line is a KWW fit of the $\alpha$ relaxation
($\beta\!=\!0.72$).}\normalsize}\label{fig1}
\end{figure}

It is well established for the $\alpha$ relaxation of polymer melts that the time
dependence differs from a simple exponential behavior and the temperature dependence does
not obey an Arrhenius law.\cite{Ed_96} Therefore, we first demonstrate that the studied
PEO and PPO models show these key features. While the incoherent intermediate scattering
function
\begin{equation}\label{eq_sqt}
F_s(q,t)=\langle\cos\{\mathbf{q}\cdot[\mathbf{r}(\tilde{t}_0\!+\!t)\!-\!\mathbf{r}(\tilde{t}_0)]\}\rangle
\end{equation}
provides us with information about translational motion, the orientational
autocorrelation function
\begin{equation}\label{eq_f2}
F_2(t)=\frac{1}{2}\,\langle\,3\,[\mathbf{e}(\tilde{t}_0\!+\!t)\cdot\mathbf{e}(\tilde{t}_0)]^2-1\rangle
\end{equation}
yields insights into rotational motion. Specifically, the scattering function $F_s$
depends on the translational displacements
$[\mathbf{r}(\tilde{t}_0\!+\!t)\!-\!\mathbf{r}(\tilde{t}_0)]$ of the atoms during the
time interval $t$, where the absolute value of the scattering vector,
$q\!=\!|\mathbf{q}|$, determines the length scale on which dynamics is probed. Throughout
this contribution, we focus on the translational motion of the oxygen atoms, but we
ensured that qualitatively similar findings are obtained for the carbon atoms. The
orientational correlation function $F_2$ depends on the angular displacements
$|\mathbf{e}(\tilde{t}_0\!+\!t)\cdot\mathbf{e}(\tilde{t}_0)|$ during the time interval
$t$. We study the reorientation of the C-H bonds and $\mathbf{e}(\tilde{t}\,)$ is the
unit vector describing the direction of a C-H bond at time $\tilde{t}$. $^2$H NMR
stimulated-echo experiments were used to measure $F_2(t)$ for C-D bonds of deuterated
PPO.\cite{Vo_06} Mimicking the experimental situation, we restrict the analysis to C-H
bonds in the methylene groups to avoid effects from fast threefold methyl group jumps.
Finally, in Eqs.\ (\ref{eq_sqt}) and (\ref{eq_f2}), the brackets $\langle\dots\rangle$
denote the average over various time origins $\tilde{t}_0$ and over all atoms or bonds
belonging to the considered atomic or bond species.

\begin{figure}
\centering
\includegraphics[width=8.0cm]{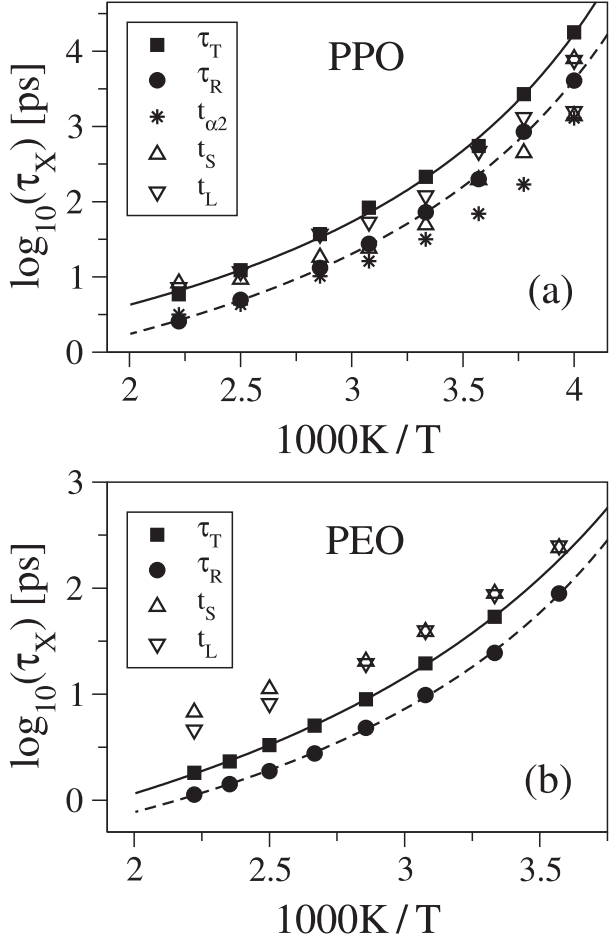}
\caption{\small \textsl{Time constants characterizing the dynamics of the (a) PPO and (b)
PEO models. The non-Gaussian parameter $\alpha_2$, the mean cluster size $S_w$ and the
mean string length $L_w$ are a maximum at $t_{\alpha 2}$, $t_{S}$ and $t_{L}$,
respectively. Moreover, $F_s(q,\tau_{T})\!=\!1/e$ and $F_2(\tau_{R})\!=\!1/e$. The solid
and dashed lines are VFT interpolations of $\tau_{T}(T)$ (PPO: $T_0\!=\!179\mathrm{\,K}$,
PEO: $T_0\!=\!179\mathrm{\,K}$) and $\tau_{R}(T)$ (PPO: $T_0\!=\!175\mathrm{\,K}$, PEO:
$T_0\!=\!191\mathrm{\,K}$).}\normalsize}\label{fig2}
\end{figure}

First, we use these correlation functions to ascertain the temperature dependence of the
translational and rotational motion associated with the $\alpha$ relaxation. To address
the translational aspect, we calculate $F_s(q,t)$ using $q\!=\!1.31\,\mathrm{\AA}^{-1}$
and $q\!=\!0.91\,\mathrm{\AA}^{-1}$ for PEO and PPO, respectively. These values of the
momentum transfer correspond to the respective position of the first maximum of the
intermolecular oxygen-oxygen pair distribution functions.\cite{Ah_00,Sm_96} In Fig.\
\ref{fig1}, we see that the incoherent intermediate scattering functions $F_s(q,t)$ for
the oxygen atoms of PPO show a pronounced temperature dependence, in particular at low
temperatures. To quantify the slowdown of the structural relaxation, we extract
temperature dependent translational and rotational correlation times according to
$F_s(q,\tau_T)\!=\!1/e$ and $F_2(\tau_{R})\!=\!1/e$. Figure \ref{fig2} shows the results
for PEO and PPO. For both models, $\tau_T$ and $\tau_R$ exhibit a comparable temperature
dependence that cannot be described by an Arrhenius law, as expected for polymer melts.
Rather, a Vogel-Fulcher-Tammann (VFT) law,\cite{VFT}
\begin{equation}\label{eq_vft}
\tau(T)=\tau_\infty\exp\left(\frac{B}{T\!-\!T_0}\right),
\end{equation}
enables good interpolations of the data. VFT fits to $\tau_T$ ($\tau_R$) yield
$T_0\!=\!179\mathrm{\,K}$ ($T_0\!=\!191\mathrm{\,K}$) for PEO and
$T_0\!=\!179\mathrm{\,K}$ ($T_0\!=\!175\mathrm{\,K}$) for PPO. The results for PPO are in
reasonable agreement with $T_0\!=\!157\!-\!170\mathrm{\,K}$ obtained in light scattering
and dielectric spectroscopy studies.\cite{Be_97,Le_99}

Inspecting the shape of the scattering functions in Fig.\ \ref{fig1}, we see
nonexponential decays. As expected for glass-forming liquids,\cite{Ed_96} the time
dependence in the $\alpha$-relaxation regime is well described by a
Kohlrausch-Williams-Watts (KWW) function, or, equivalently, stretched exponential
function
\begin{equation}\label{eq_kww}
F_{KWW}(t)=A\,\exp\left[-\left(\frac{t}{\tau}\right)^{\beta}\right]\;\;\;\;(0\!\leq\!\beta\!\leq\!1).
\end{equation}
Here, $\tau$ and $\beta$ quantify the time scale and the stretching, respectively.
Fitting $F_s(q,t)$ to a KWW function, we find stretching parameters
$\beta\!\approx\!0.63$ for PEO and $\beta\!\approx\!0.71$ for PPO, independent of
temperature in the studied temperature ranges. For PPO, the time-temperature
superposition is further demonstrated in the inset of Fig.\ \ref{fig1}. Clearly, the
curves coincide in the $\alpha$-relaxation regime when the time axis is scaled with the
correlation time $\tau$. KWW interpolations of $F_2(t)$ yield stretching parameters
$\beta\!\approx\!0.36$ for PEO and $\beta\!\approx\!0.48$ for PPO, consistent with
stretching parameters obtained from $^2$H NMR stimulated-echo experiments near
$T_g$.\cite{Vo_06} Thus, with respect to both temperature and time dependence, the
structural relaxation of the studied polymer models resembles that of PEO and PPO melts,
confirming the quality of the used force fields.\cite{Bo_03_2}

We note that, for PPO at $T\!=\!250\mathrm{\,K}$, the translational and rotational
correlation functions give evidence for some deviations from time-temperature
superposition in the early stages of the decay, where the curves decrease significantly
faster than expected from the KWW behavior, see Fig.\ \ref{fig1}. Since dielectric
spectroscopy studies demonstrated that PPO exhibits a Johari-Goldstein secondary
relaxation, which starts to separate from the primary relaxation at
$T\!\approx\!250\mathrm{\,K}$, \cite{Be_97,Si_91,Le_99} we attribute these deviations to
the onset of this secondary process.

\subsection{MCT analysis}\label{sec_mct}

\begin{figure}
\centering
\includegraphics[width=8.0cm]{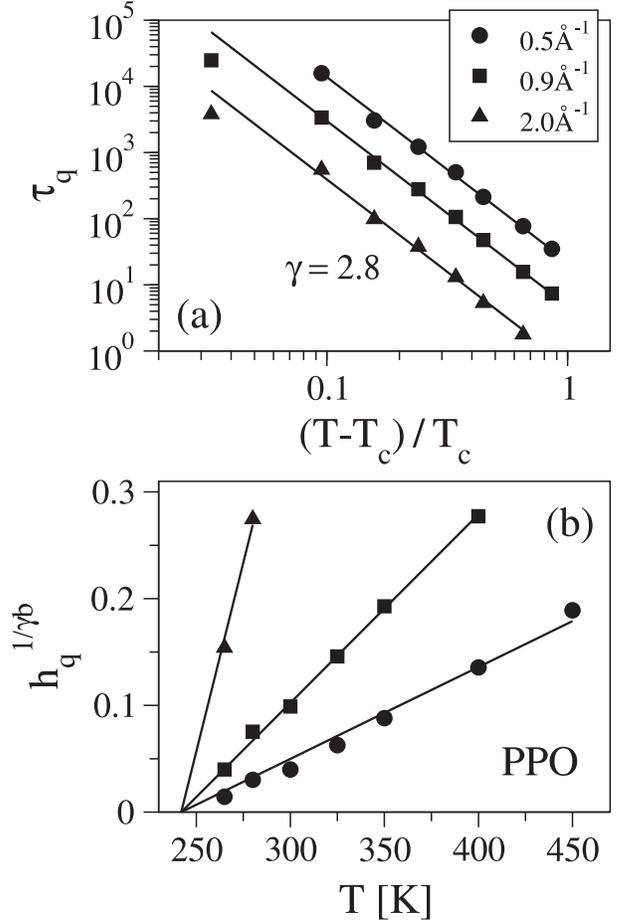}
\caption{\small \textsl{Test of MCT predictions for PPO on the basis of the incoherent
intermediate scattering functions $F_s(q,t)$ of the oxygen atoms. (a) Correlation times
$\tau_{q}$ resulting from KWW fits of the $\alpha$-relaxation regime for the indicated
values of $q$. The solid lines are MCT power laws
$\tau_q\!\propto\!(T\!-\!T_c)^{-\gamma}$ where $T_c\!=\!242\mathrm{\,K}$ and
$\gamma\!=\!2.8$. (b) Temperature dependence of $h_q^{1/\gamma b}$, using $\gamma
b\!=\!1.44$  for $q\!=\!0.50\mathrm{\,\AA^{-1}}$, $\gamma b\!=\!1.34$ for
$q\!=\!0.91\mathrm{\,\AA^{-1}}$, and $\gamma b\!=\!1.16$ for
$q\!=\!2.00\mathrm{\,\AA^{-1}}$. The solid lines are linear interpolations $h_q^{1/\gamma
b}\!\propto\!(T\!-\!242\mathrm{\,K})$.}\normalsize}\label{fig3}
\end{figure}

The time temperature superposition observed for the $\alpha$ relaxation of the PEO and
PPO models is in agreement with MCT.\cite{Go_92} Therefore, we now check whether further
predictions of this theory are fulfilled. According to MCT, the particles of
glass-forming liquids are trapped in cages formed by their neighbors for some time until
an escape from these cages is possible during the structural relaxation. As a consequence
of this trapping, a plateau regime, or, equivalently, $\beta$-relaxation regime,
preceding the $\alpha$-relaxation regime develops when the temperature is decreased
towards the critical temperature $T_c$. Quantitatively, MCT predicts a power-law
divergence of the $\alpha$-relaxation time at the critical temperature $T_c$:
\begin{equation}\label{eq_mct}
\tau(T)\propto(T-T_c)^{-\gamma}.
\end{equation}
Another key prediction of MCT is the factorization theorem for the $\beta$-relaxation
regime. It states that, for such times, all correlation functions, in particular, the
incoherent intermediate scattering functions for different values of the momentum
transfer, can be written as
\begin{equation}\label{eq_fac}
F_s(q,t)=f_q^c+a_qG(t).
\end{equation}
Here, the plateau value $f_q^c$ and the amplitude $a_q$ depend on the value of $q$, while
the $\beta$ correlator $G(t)$ is independent of the observable. If the factorization
theorem is obeyed, the curves
\begin{equation}\label{eq_r}
R(q,t)=\frac{F_s(q,t)-F_s(q,t^{\prime})}{F_s(q,t^{\prime\prime})-F_s(q,t^{\prime})}
\end{equation}
for different values of the momentum transfer will collapse onto a master curve, provided
the times $t^\prime$ and $t^{\prime\prime}$ are chosen inside the $\beta$-relaxation
regime.\cite{Gl_00} Hence, calculation of $R(q,t)$ allows one to check the factorization
theorem without invoking a fitting procedure. $G(t)$ can be expanded for times close to
the central $\beta$-relaxation time $t_\sigma$. The expansion for $t\!>\!t_\sigma$ leads
to the von Schweidler law
\begin{equation}\label{eq_von}
F_s(q,t)=f_q^{c}-h_qt^b,
\end{equation}
showing that a power law characterized by the universal von Schweidler exponent $b$
describes the initial stages of the decay from the plateau. For temperatures $T\!>\!T_c$,
the amplitude $h_q$ decreases upon cooling according to
\begin{equation}\label{eq_hq}
h_q(T)\propto(T-T_c)^{\gamma b}.
\end{equation}
Finally, within MCT, $b$ and $\gamma$ are related via the exponent parameter $\lambda$:
\begin{equation}\label{eq_exp}
\lambda=\frac{\Gamma^2(1-a)}{\Gamma(1-2a)}=\frac{\Gamma^2(1+b)}{\Gamma(1+2b)}\;,\;\;\;\;\;
\gamma=\frac{1}{2a}+\frac{1}{2b}\,.
\end{equation}

To check these MCT predictions for the PPO model, we analyze the scattering functions
$F_s(q,t)$ of the oxygen atoms for various values of the momentum transfer $q$. First, we
fit a KWW function to the scattering functions in the $\alpha$-relaxation regime. Figure
\ref{fig3}(a) shows the temperature dependent correlation times $\tau_q$ resulting from
these fits for three values of $q$. A MCT power law, see Eq.\ (\ref{eq_mct}), with
$T_c\!=\!242\!\pm\!2\mathrm{\,K}$ and $\gamma\!=\!2.8\!\pm\!0.1$ nicely describes all
data at $T/T_c\!-\!1\!\geq\!0.1$. These findings are in reasonable agrement with
$T_c\!=\!236\!\pm\!2\mathrm{\,K}$ and $\gamma\!=\!3.7\!\pm\!0.8$ from experimental
studies.\cite{Si_92,Be_97} However, there are deviations from the MCT power law at the
lowest temperature $T/T_c\!-\!1\!\approx\!0.03$. At the present, it is not clear whether
these deviations are related to the possible onset of the Johari-Goldstein secondary
relaxation near $T_c$. Qualitatively similar deviations from the MCT predictions were
reported for a chemically realistic model of polybutadiene in the vicinity of the
critical temperature.\cite{Pa_06}

\begin{figure}
\centering
\includegraphics[width=8.25cm]{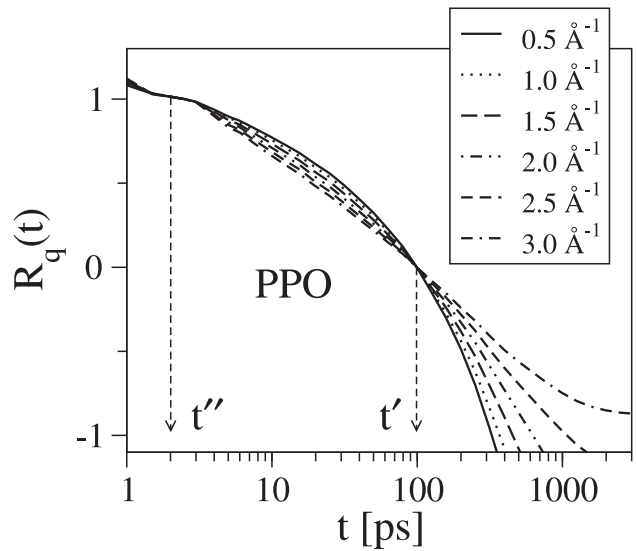}
\caption{\small \textsl{Test of the MCT factorization theorem for PPO at
$T\!=\!265\mathrm{\,K}$. The functions $R(q,t)$ were obtained from the data $F_s(q,t)$
for the oxygen atoms according to Eq.\ (\ref{eq_r}). The values of the momentum transfer
$q$ are indicated and times $t^{\prime\prime}\!=\!2\mathrm{\,ps}$ and
$t^{\prime}\!=\!100\mathrm{\,ps}$ were used.}\normalsize}\label{fig4}
\end{figure}

Next, we fit the von Schweidler law to the scattering functions $F_s(q,t)$. Due to a
possible interference of the Johari-Goldstein secondary relaxation, we exclude the data
for the lowest temperature $T\!=\!250\!\mathrm{\,K}$ from further MCT analysis. The von
Schweidler law enables a good interpolation of the decays in the late-$\beta$/
early-$\alpha$ relaxation regime, see Fig.\ \ref{fig1}. For
$q\!=\!0.91\,\mathrm{\AA}^{-1}$, the von Schweidler fits yield $f_q
^c\!=\!0.93\!\pm\!0.01$ and $b\!=\!0.46\!\pm\!0.01$ independent of temperature. According
to Eq.\ (\ref{eq_exp}), this value of the von Schweidler exponent $b$ translates into
$\gamma\!=\!2.9$, in agreement with $\gamma\!=\!2.8\!\pm\!0.1$ determined from the
temperature dependent correlation times $\tau_q$. To check the validity of Eq.\
(\ref{eq_hq}), we use the exponents $b\!=\!0.46$ and $\gamma\!=\!2.9$ and plot the
temperature dependence of $h_q^{1/\gamma b}$ in Fig.\ \ref{fig3}(b). In harmony with the
MCT prediction, there is a linear relationship and, by extrapolation, $h_q^{1/\gamma b}$
vanishes at the critical temperature $T_c\!=\!242\mathrm{\,K}$, extracted from the
temperature dependence of the $\alpha$ relaxation.

However, the MCT predictions are not fulfilled when including the scattering functions
for other values of $q$ into the analysis. Von Schweidler fits for
$q\!=\!0.5\,\mathrm{\AA}^{-1}$ and $q\!=\!2.0\,\mathrm{\AA}^{-1}$ yield
$b\!=\!0.59\!\pm\!0.02$ and $b\!=\!0.23\!\pm\!0.02$, respectively, corresponding to
$\gamma\!=\!2.4$ and $\gamma\!=\!5.0$. Thus, there is no universal von Schweidler
exponent and, for high and low values of $q$, the values of $\gamma$ calculated from the
von Schweidler exponents deviate from $\gamma\!=\!2.8$ resulting from the temperature
dependence of the $\alpha$ relaxation. To demonstrate the violation of the factorization
theorem independent of any fitting routine, we show $R(q,t)$ for different values $q$ in
Fig.\ \ref{fig4}. The data was calculated from $F_s(q,t)$ for the oxygen atoms of PPO at
$T\!=\!265\mathrm{\,K}$. Clearly, the data do not collapse onto a master curve in the
$\beta$-relaxation regime, indicating the violation of the factorization theorem.
Qualitatively similar results were observed at all studied temperatures.

\subsection{Spatially heterogeneous dynamics}\label{sec_shd}

\begin{figure}
\centering
\includegraphics[width=8.5cm]{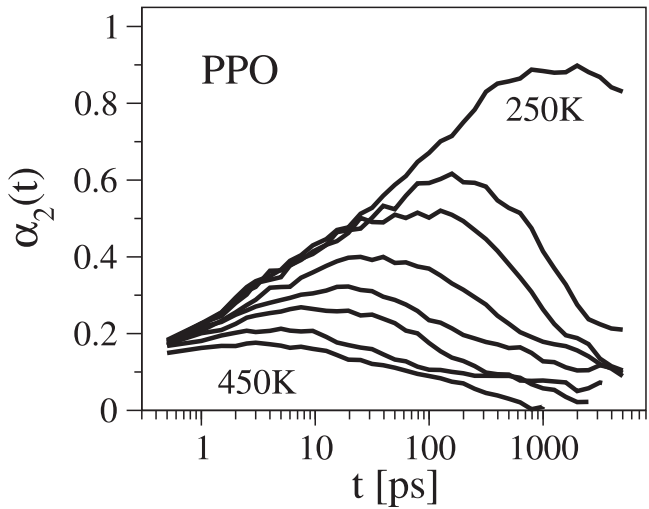}
\caption{\small \textsl{Non-Gaussian parameter $\alpha_2(t)$ for the oxygen atoms of PPO
at various temperatures ($450\mathrm{\,K}$, $400\mathrm{\,K}$, $350\mathrm{\,K}$,
$325\mathrm{\,K}$, $300\mathrm{\,K}$, $280\mathrm{\,K}$, $265\mathrm{\,K}$,
$250\mathrm{\,K}$).}\normalsize}\label{fig5}
\end{figure}

Other models of the glass transition focus on the heterogeneity and the cooperativity of
the dynamics.\cite{Ad_65,Ga_02} The importance of these effects was demonstrated in MD
simulation studies on various glass-forming
liquids.\cite{Ko_97,Dol_98,Don_99_1,Don_99_2,Be_99,Gl_00_2,Ge_01,Ge_04,Don_98,Ai_03,Qi_99,Gi_03,Vo_04_1,Vo_04_2,Te_04}
Here, we investigate the relevance of heterogeneity and cooperativity for the first time
for chemically realistic polymer models. For various model-glass formers, it was found
that the distribution of scalar particle displacements
$|\mathbf{r}(\tilde{t}_0\!+\!t)\!-\!\mathbf{r}(\tilde{t}_0)|$ deviates from a Gaussian at
intermediate times $t$ between ballistic and diffusive motion. These deviations, which
are a first indication for the existence of heterogeneous dynamics, can be quantified by
the non-Gaussian parameter
\begin{equation}\label{eq_a2}
\alpha_2(t)=\frac{3}{5}\,\frac{\langle[\mathbf{r}(\tilde{t}_0\!+\!t)\!-\!\mathbf{r}(\tilde{t}_0)]^4\rangle}{\langle[\mathbf{r}(\tilde{t}_0\!+\!t)\!-\!\mathbf{r}(\tilde{t}_0)]^2\rangle^2}-1.
\end{equation}
Figure \ref{fig5} shows $\alpha_2(t)$ for the oxygen atoms of PPO at various
temperatures. We see that the non-Gaussian parameter exhibits a maximum in the
late-$\beta$/ early-$\alpha$ relaxation regime, consistent with findings for various
model-glass formers.\cite{Ko_97,Qi_99,Ai_03,Gi_03,Vo_04_2} When the temperature is
decreased, the position of the maximum, $t_{\alpha 2}$, shifts to longer times. In Fig.\
\ref{fig2}(a), we see that the temperature dependence of $t_{\alpha 2}$ is somewhat
weaker, but still comparable to that of the $\alpha$-relaxation time. The maximum
$\alpha_2(t_{\alpha 2})$ increases upon cooling, in particular near the critical
temperature $T_c$, but the values are relatively small. More precisely, at comparable
temperatures near $T_c$, $\alpha_2(t_{\alpha 2})\!=\!1.3\!-\!2.5$ was reported for the
other studied models of glass-forming liquids, except for the silicon atoms in silica
showing $\alpha_2(t_{\alpha 2})\!=\!0.8$, see Ref.\ \onlinecite{Vo_04_2} for a detailed
comparison. For PEO, the non-Gaussian parameter is even smaller than for PPO. This
results in some ambiguities when extracting the maximum positions so that we refrain from
discussing the temperature dependence of $t_{\alpha 2}$ for PEO.

To ascertain the spatial heterogeneity of the PEO and PPO dynamics, we demonstrate that
highly mobile oxygen atoms form clusters larger than expected from random statistics.
Following previous studies,\cite{Don_99_2,Ge_01,Ge_04,Gi_03,Vo_04_2} we characterize the
particle mobility in a time interval $t$ by the scalar displacement and select the 5\%
most mobile oxygen atoms for further analysis. Then, we define a cluster as a group of
the most mobile oxygen atoms that reside in the first neighbor shells of each other. For
both polymers, we use the position of the first minimum of the respective intermolecular
oxygen-oxygen pair distribution function as criterion for the extension of the neighbor
shell. Based on the probability distribution $p_s(n,t)$ of finding a cluster of size $n$
for a time interval $t$, we calculate the weight-averaged mean cluster size
\begin{equation}\label{eq_sw}
S_w(t)=\frac{\sum_n n^2 p_s(n,t)}{\sum_n n\, p_s(n,t)}\,.
\end{equation}
This quantity measures the average size of a cluster to which one of the most mobile
oxygen atoms belongs. Previously, it was shown that the conclusions resulting from such
analysis are unchanged when the fraction of highly mobile particles is varied in a
meaningful range.\cite{Vo_04_2}

\begin{figure}
\centering
\includegraphics[width=8.0cm]{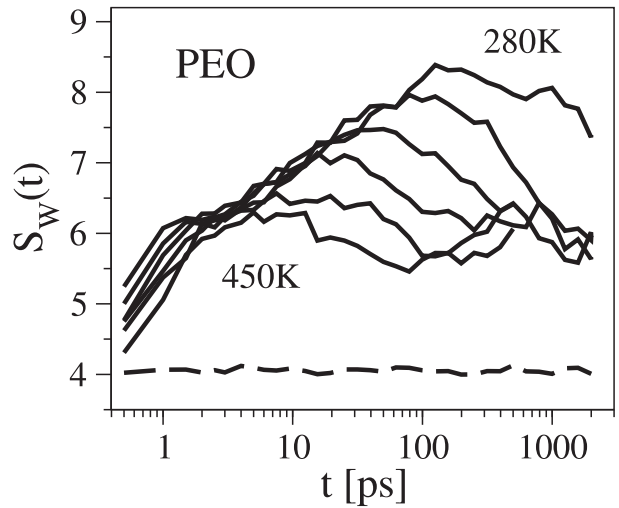}
\caption{\small \textsl{Mean cluster size $S_w(t)$ of highly mobile oxygen atoms for PEO
at temperatures $450\mathrm{\,K}$, $400\mathrm{\,K}$, $350\mathrm{\,K}$,
$325\mathrm{\,K}$, $300\mathrm{\,K}$, and $280\mathrm{\,K}$. When the particles used for
the analysis are chosen irrespective of their mobilities, a mean cluster size of
$S_w\!\approx\!4$ results, as indicated by the dashed line. }\normalsize}\label{fig6}
\end{figure}

\begin{figure}
\centering
\includegraphics[width=8.25cm]{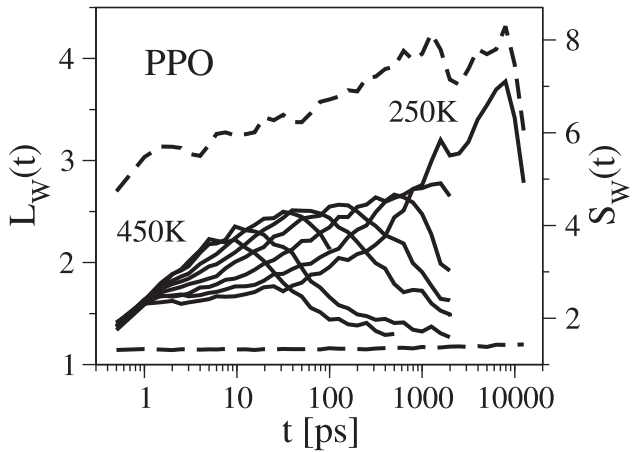}
\caption{\small \textsl{Mean string length $L_w(t)$ of highly mobile oxygen atoms for PPO
at temperatures $450\mathrm{\,K}$, $400\mathrm{\,K}$, $350\mathrm{\,K}$,
$325\mathrm{\,K}$, $300\mathrm{\,K}$, $280\mathrm{\,K}$, $265\mathrm{\,K}$, and
$250\mathrm{\,K}$. When the oxygen atoms used for the analysis are chosen irrespective of
their mobilities, a mean string length of $L_w\!\approx\!1.1$ results, as indicated by
the bottommost dashed line. The upmost dashed line is the mean cluster size $S_w(t)$ of
highly mobile oxygen atoms for PPO at $T\!=\!250\mathrm{\,K}$.}\normalsize}\label{fig7}
\end{figure}

The mean cluster size $S_w(t)$ is displayed for PEO at various temperatures in Fig.\
\ref{fig6}. It is evident that $S_w(t)$ shows a maximum, which increases upon cooling. By
contrast, a time and temperature independent small size $S_w\!\approx\!4$ characterizes
the clusters for the case of random statistics, i.e., when 5\% of the oxygen atoms are
chosen for analysis irrespective of their mobilities. These results clearly demonstrate
the existence of spatially heterogeneous dynamics. The transient nature of the clusters
can be quantified, when we determine the times $t_S$ at which $S_w(t)$ is a maximum. In
Fig.\ \ref{fig2}, we see for PEO and PPO that the clusters are largest in the
$\alpha$-relaxation regime at all studied temperatures. For PPO, we observe that the
single maximum of $S_w(t)$ at high temperatures splits into two peaks near $T_c$, see
Fig.\ \ref{fig7}. This can be taken as another hint that a primary and a secondary
relaxation coexist at sufficiently low temperatures.

In Fig.\ \ref{fig6}, a closer inspection of the data for the highest temperatures reveals
that $S_w(t)$ increases at long times in the diffusive regime. This increase is the mere
consequence of the chain connectivity. At sufficiently long times, the displacements of
the individual atoms are largely determined by the displacement of the center-of-mass of
the respective polymer chain and, hence, those oxygen atoms are highly mobile that belong
to chains, showing the largest center-of-mass displacements from the statistical
distribution. Due to their spatial proximity along the chain, these atoms form extended
clusters so that $S_w(t)$ increases when the single particle displacements start to
become dominated by the center-of-mass displacements. By contrast, the maximum at shorter
times is not due to chain connectivity.

It has been shown that cooperative string-like motion is an important channel for the
relaxation of highly mobile particles in simple model-glass
formers.\cite{Don_98,Don_99_2,Ai_03,Ge_04} Therefore, we investigate the relevance of
string-like motion for the chemically realistic PEO and PPO models. Following these
previous studies, we construct strings by connecting any two oxygen atoms $i$ and $j$ if
the condition
\begin{displaymath}\label{eq_lw}
\mathrm{min}[\,|\mathbf{r}_i(\tilde{t}_0)\!-\!\mathbf{r}_j(\tilde{t}_0\!+\!t)|,|\mathbf{r}_j(\tilde{t}_0)\!-\!\mathbf{r}_i(\tilde{t}_0\!+\!t)|\,]<\delta
\end{displaymath}
holds for the atomic positions at two different times and set $\delta$ to about 55\% of
the intermolecular oxygen-oxygen distance. Then, this condition means that one oxygen
atom has moved and an another oxygen atom has occupied its position. We checked that our
conclusions are not altered when $\delta$ is varied in a meaningful range. Using the
above criterion, we determine the probability $p_l(l,t)$ of finding a string of length
$l$ for a time interval $t$ and calculate the weight-averaged mean string length $L_w(t)$
in analogy with Eq.\ (\ref{eq_sw}).

Figure \ref{fig7} shows $L_w(t)$ for PPO at various temperatures. Evidently, the strings
grow and shrink in time and they are substantially longer than that resulting from random
statistics. Upon cooling, the position of the maximum, $t_L$, shifts to longer times and
the height of the maximum increases, where the growth is particularly prominent in the
vicinity of $T_c$. Thus, string-like motion is an important phenomenon at sufficiently
low temperatures. $L_w(t)$ and $S_w(t)$ show a similar behavior. In particular, both
quantities exhibit a two-peak signature at $T\!=\!250\mathrm{\,K}$. The temperature
dependence of $t_L$ and $t_S$ is compared in Fig.\ \ref{fig2}. For PEO and PPO, the
strings, like the clusters, are largest in the $\alpha$-relaxation regime. By contrast,
simulation studies on simple glass-forming liquids, including a bead-spring polymer
model, found that $t_L$ and $t_S$ are located in the late-$\beta$/ early-$\alpha$
relaxation regime, i.e., at significantly shorter
times.\cite{Don_98,Don_99_2,Ge_04,Gi_03,Vo_04_2,Ge_01,Ai_03} We conclude that the PEO and
PPO models do show spatially heterogenous and cooperative dynamics, however the
characteristics of these effects differ from that for simple model-glass formers.

\subsection{Conformational relaxation}\label{sec_con}

\begin{figure}
\centering
\includegraphics[width=8.25cm]{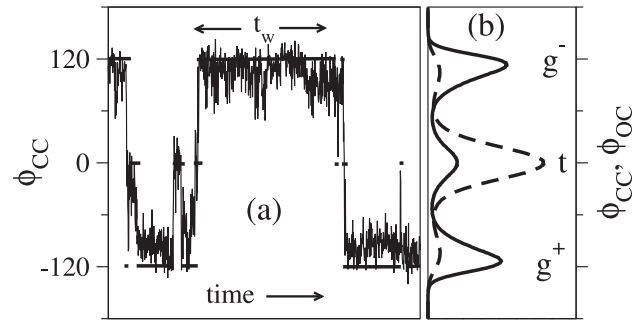}
\caption{\small \textsl{ Results for PEO at $T\!=\!280\mathrm{\,K}$. (a) Typical
trajectory $\phi_{cc}(\tilde{t}\,)$ of an OCCO dihedral angle during a time interval of
$1.15\mathrm{\,ns}$. The straight lines mark the discontinuous trajectory
$s_{cc}(\tilde{t}\,)$ resulting from mapping of the trajectory $\phi_{cc}(\tilde{t}\,)$
onto the conformational states $g^-$, $t$, and $g^+$. (b) Probability distributions of
the OCCO (solid line) and COCC (dashed line) dihedral angles, $p\,(\phi_{cc})$ and
$p\,(\phi_{oc})$, respectively. }\normalsize}\label{fig9}
\end{figure}

Local conformational dynamics are of central importance for the structural relaxation of
polymers.\cite{Pa_04} In the following, we investigate the torsional motion of the PEO
model by analyzing the angular trajectories $\phi_{cc}(\tilde{t}\,)$ and
$\phi_{oc}(\tilde{t}\,)$, describing the time evolution of the OCCO and COCC dihedral
angles, respectively. Figure \ref{fig9} presents a typical trajectory
$\phi_{cc}(\tilde{t}\,)$. Comparison with the probability distribution of the dihedral
angle, $p\,(\phi_{cc})$, shows that the torsional motion is comprised of well defined
transitions between the $gauche^-$ ($g^-$), $trans$ ($t$), and $gauche^+$ ($g^+$) states,
suggesting that a discretization is useful for an analysis of the conformational
dynamics. Therefore, we map the trajectories $\phi_{x}(\tilde{t}\,)$ ($x\!=\!cc,oc$) onto
discrete sequences $s_{x}(\tilde{t}\,)$ of the conformational states, see Fig.\
\ref{fig9}. In this way, we eliminate effects from librational motions, which do not lead
to structural relaxation. Then, analysis of the discrete sequences $s_{x}(\tilde{t}\,)$
enables a straightforward characterization of the relevant torsional motion, e.g., in
terms of waiting times $t_w$ and back-jump probabilities $p_{\,b}$, see below.

\begin{figure}
\centering
\includegraphics[width=8.25cm]{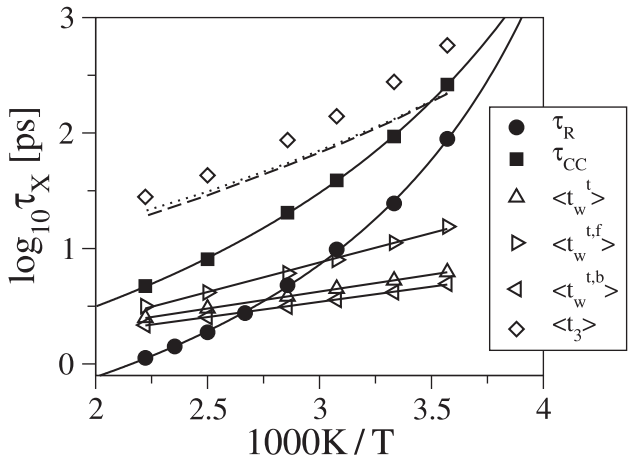}
\caption{\small \textsl{Time constants characterizing the conformational dynamics of the
OCCO dihedrals in the PEO model. We show the torsional correlation times $\tau_{cc}$,
defined as $P_{cc}(\tau_{cc})=\!1/e$, and the rotational correlation times $\tau_{R}$,
see Fig.\ \ref{fig2}, together with VFT interpolations. Moreover, the mean waiting times
of the OCCO dihedrals in the $t$ state are displayed. While all escape processes from the
$t$ state are considered for the calculation of $\langle t_w^t \rangle$, $\langle
t_w^{t,f} \rangle$ ($\langle t_w^{t,b} \rangle$) characterizes explicitly the forward
(backward) jumps $g^\pm\!\rightarrow\!t\!\rightarrow\!g^\mp$
($g^\pm\!\rightarrow\!t\!\rightarrow\!g^\pm$). The solid lines are Arrhenius fits
($\langle t_w^t \rangle$: $E_a\!=\!0.025\mathrm{\,eV}$, $\langle t_w^{t,f} \rangle$:
$E_a\!=\!0.044\mathrm{\,eV}$, $\langle t_w^{t,b} \rangle$: $E_a\!=\!0.023\mathrm{\,eV}$).
Finally, we present the mean times $\langle t_3 \rangle$ needed for an OCCO dihedral to
visit each of the three conformational states at least once. Calculated values of
$\langle t_3 \rangle$ are shown as dotted and dashed lines. While
$g^\pm\!\leftrightarrow\!g^\mp$ transitions were neglected when calculating the former
data, these transitions were taken into account when computing the latter data, see text
for details.}\normalsize}\label{fig8}
\end{figure}

To study the time scale of the conformational relaxation, we use the discrete sequences
$s_x$ and determine the probabilities $p_{x}(t)$ of finding a dihedral in the same
conformational state at two times separated by a time interval $t$. Due to the finite
number of conformational states, these probabilities exhibit a finite and temperature
dependent plateau value $p_{x}(\infty)$. To remove this effect, we calculate the
torsional correlation functions
\begin{equation}\label{eq_p}
P_{x}(t)=\frac{\langle p_{x}(t)\rangle-\langle p_{x}(\infty)\rangle}{1-\langle
p_{x}(\infty)\rangle}
\end{equation}
and determine torsional correlation times $\tau_{x}$ according to
$P_{x}(\tau_{x})\!=\!1/e$. In Fig.\ \ref{fig8}, we compare the torsional correlation time
$\tau_{cc}$ with the rotational correlation time $\tau_R$, characterizing the
reorientation of the C-H bond vectors. We see that both time constants show a comparable
temperature dependence, implying that the conformational and the structural relaxations
are related, consistent with results for other polymer models.\cite{Pa_04,Sm_97} Some
deviations in the temperature dependence are expected since, in particular at high
temperatures, librational motions affect the rotational correlation times more than the
torsional correlation times obtained from analysis of the discrete jump sequences.
Moreover, the conformational dynamics of several dihedral species render the orientation
of a given C-H bond time dependent and, hence, the absolute values of the rotational and
torsional correlation times should differ.

Next, we study the waiting times $t_w$ in the conformational states, i.e., the time
intervals between two subsequent conformational transitions. We find that the waiting
times differ between the OCCO and COCC dihedral angles and, for each dihedral species,
they depend on the conformational state. Therefore, we separately determine the waiting
times for each dihedral species and conformational state. The temperature dependent mean
waiting time of the OCCO dihedrals in the $t$ state, $\langle t_w^t\rangle$, is included
in Fig.\ \ref{fig8}. We see that the mean waiting time follows an Arrhenius law with a
small activation energy $E_a\!=\!0.025\mathrm{\,eV}$. Qualitatively similar results are
observed for all other dihedral species and conformational states. Hence, the temperature
dependence of the mean waiting times is much weaker than that of the rotational and
torsional correlation times, indicating that the longer residence times in the
conformational states at lower temperatures are not sufficient to explain the strong
slowdown of the $\alpha$ relaxation upon cooling, in harmony with results for other
all-atom polymer models.\cite{Pa_04}

\begin{figure} \centering
\includegraphics[width=8.25cm]{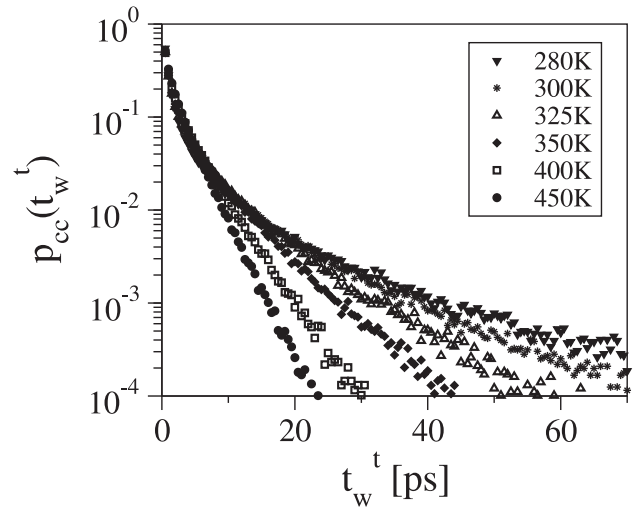}
\caption{\small \textsl{Probability distributions $p_{cc}(t_w^t)$ characterizing the
waiting times of the OCCO dihedrals in the $t$ state. Results for PEO at various
temperatures are compared.}\normalsize}\label{fig10}
\end{figure}

Closer insights into the nature of the conformational dynamics are available from the
probability distributions of the waiting times, $p_{x}(t_w)$. Figure \ref{fig10} shows
the distributions $p_{cc}(t_w^t)$, which characterize the waiting times of the OCCO
dihedrals in the $t$ state. In particular at the lower temperatures, there are
substantial deviations from an exponential waiting-time distribution, indicating that a
Markov process does not apply to the conformational dynamics. At short waiting times, we
see a nonexponential decay, which exhibits a weak temperature dependence. At long waiting
times, there is a nearly exponential and more temperature dependent decay. Qualitatively
similar waiting-time distributions are observed for all dihedral species and
conformational states, suggesting that the distributions are comprised of two
contributions governing the behavior at short and long waiting times, respectively.

\begin{figure}
\centering
\includegraphics[width=8.25cm]{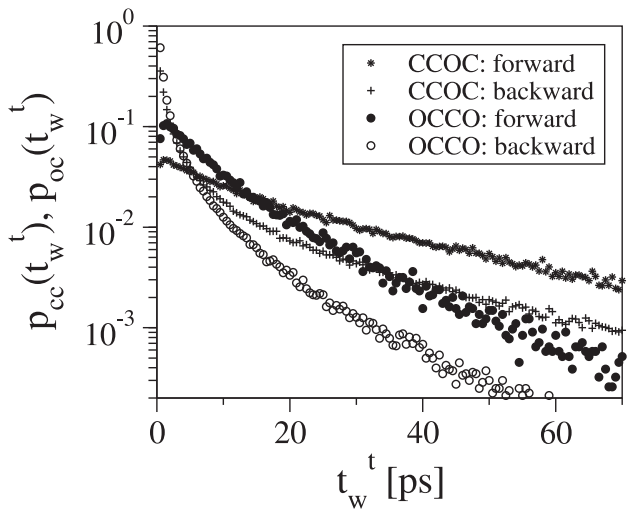}
\caption{\small \textsl{Probability distributions $p_{cc}(t_w^t)$ and $p_{oc}(t_w^t)$
characterizing the waiting times of the OCCO and COCC dihedrals in the $t$ state,
respectively. Results for PEO at $T\!=\!300\mathrm{\,K}$ are shown. For both dihedral
species, we distinguish the waiting times prior to a forward jump
($g^\pm\!\rightarrow\!t\!\rightarrow\!g^\mp$) from that prior to a backward jump
($g^\pm\!\rightarrow\!t\!\rightarrow\!g^\pm$).}\normalsize}\label{fig11}
\end{figure}

In general, one can distinguish between forward and backward jumps. Let
$\sigma_{i-1}\!\rightarrow\!\sigma_i\!\rightarrow\!\sigma_{i+1}$ be a sequence of three
subsequently visited conformational states. Then, a forward and a backward jump in the
state $\sigma_i$ are associated with $\sigma_{i-1}\!\neq\!\sigma_{i+1}$ and
$\sigma_{i-1}\!=\!\sigma_{i+1}$, respectively. Unlike backward jumps, forward jumps
enable visiting all three conformational states and, hence, they may be the cornerstone
of conformational relaxation. Since transitions $g^\pm\!\rightarrow\!g^\mp$ are rare for
the OCCO and COCC dihedrals at the studied temperatures, the behavior in the $t$ state is
of particular importance for the exploration of all conformational states. Figure
\ref{fig11} shows the probability distributions $p_{x}(t_w^t)$ characterizing the waiting
times in the $t$ state during the forward ($g^\pm\!\rightarrow\!t\!\rightarrow\!g^\mp$)
and backward ($g^\pm\!\rightarrow\!t\!\rightarrow\!g^\pm$) sequences, respectively. We
see that the results for the two dihedral species are comparable. Thus, it is not
important that the $t$ state is the majority state of $\phi_{oc}$, whereas it is the
minority state of $\phi_{cc}$, see Fig.\ \ref{fig9}. However, the waiting times strongly
depend on the direction of the jumps. For both dihedral species, the forward jumps are
characterized by nearly exponential waiting-time distributions and, hence, the propensity
to perform such transition does not depend on the jump history. The backward jumps
exhibit a more complex behavior. In a semilogarithmic representation, the slope of the
curves decreases with increasing waiting time until it becomes constant, indicative of an
exponential behavior at sufficiently long waiting times. In the latter regime, the slope
is comparable for the forward and backward jumps.

These findings imply that both uncorrelated and correlated conformational transitions
occur. In some cases, the time and the direction of a jump are independent of the
history, resulting in an exponential behavior and in similar rates of forward and
backward jumps. In other cases, the dihedrals have a high tendency to return to the
previous conformational state almost immediately, i.e., to perform a correlated backward
jump, leading to the observed nonexponentiality at short waiting times. One can imagine
that this proneness to a correlated backward jump depends on the jump history. For
example, it may become weaker when time elapses after the previous transition. Also, a
correlated backward jump may be less likely when a large number of forward-backward jumps
have already taken place, i.e., when previous ''unsuccessful'' attempts have paved the
way for a ''successful'' transition to a new state, see below.

\begin{figure}
\centering
\includegraphics[width=8.25cm]{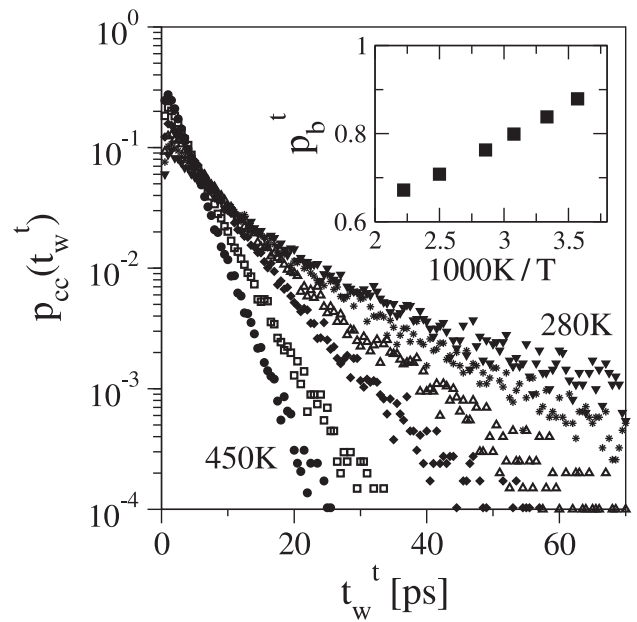}
\caption{\small \textsl{Probability distribution $p_{cc}(t_w^t)$ characterizing the
waiting times of the OCCO dihedrals in the $t$ state prior to a forward jump. Results for
PEO at various temperatures are shown. The inset displays the temperature dependence of
the probability $p_b^t$ that a forward jump $g^\pm\!\rightarrow\!t$ is followed by a
direct backward jump $t\!\rightarrow\!g^\pm$ to the initial dihedral
state.}\normalsize}\label{fig12}
\end{figure}

To further investigate the role of the forward jumps for the conformational and
structural relaxations, we analyze their temperature dependent behavior. Figure
\ref{fig12} shows the distributions $p_{cc}(t_w^t)$ characterizing the waiting times of
the OCCO dihedrals in the $t$ state prior to a forward jump at various temperatures.
Evidently, all waiting-time distributions are nearly exponential, confirming that the
point in time of a forward jump is independent of the history. In Fig.\ \ref{fig8}, we
compare the temperature dependent mean waiting times $\langle t_w^{f,t}\rangle$ and
$\langle t_w^{b,t}\rangle$ characterizing the forward and the backward jumps in the $t$
state, respectively. We see that the forward jumps exhibit a higher temperature
dependence than the backward jumps. Specifically, the activation energies amount to
$E_a\!=\!0.044\mathrm{\,eV}$ for the former and $E_a\!=\!0.023\mathrm{\,eV}$ for the
latter. However, the temperature dependence of $\langle t_w^{f,t} \rangle$ is still much
weaker than that of the torsional and rotational correlation times, corroborating our
previous conclusion that the longer waiting times at lower temperatures do not provide an
explanation for the slowdown of the $\alpha$ relaxation upon cooling.

These findings show that it is not sufficient to study single events, but it is necessary
to correlate the times and the directions of two or more conformational transitions.
First, we analyze the directions of two consecutive jumps. Specifically, we determine the
back-jump probability $p_b^{\,t}$ in the $t$ state, i.e., the probability that a
transition $g^\pm\!\rightarrow\!t$ is followed by a transition $t\!\rightarrow\!g^\pm$.
In Fig.\ \ref{fig12}, the temperature dependent back-jump probabilities $p_b^{\,t}$ of
the OCCO dihedrals are presented. We see a substantial increase of $p_b^{\,t}$ upon
cooling. Thus, when the temperature is decreased, both longer waiting times and higher
back-jump probabilities contribute to the slower exploration of the conformational
states. In other words, the increase of the back-jump probabilities leads to an
additional delay of the conformational relaxation and, hence, to deviations from an
Arrhenius behavior.

To study the exploration of the dihedral states in more detail, we determine the times
$t_3$ needed for the dihedrals to visit all three conformational states. Specifically, we
start from a randomly chosen time origin and define $t_3$ as the time elapsing until a
given dihedral has first visited each of the three conformational states at least once.
In Fig.\ \ref{fig9}, we include the mean times $\langle t_3 \rangle$ resulting from the
conformational dynamics of the OCCO dihedrals at various temperatures. Unlike the mean
waiting times, the exploration times $\langle t_3 \rangle$ do not follow an Arrhenius
law. However, their temperature dependence is still weaker than that of the torsional
correlation times, implying that the relation between the exploration process and the
conformational relaxation is not straightforward, but subtle. We note that previous work
on polybutadiene did not relate the exploration process to the primary, but rather to the
secondary relaxation.\cite{Sm_07}

The question arises whether knowledge of the back-jump probabilities and of the mean
waiting times prior to forward and backward jumps is sufficient to calculate the
exploration times. To tackle this question, we assume that, at the randomly chosen time
origin, an OCCO dihedral occupies one of the $gauche$ states, being the majority states,
and calculate the average time needed for the dihedral to visit each conformational state
at least once. First, we neglect direct transitions between the $gauche$ states due to
their rareness. Then, a forward jump in the $t$ state is necessary to visit all dihedral
states. Prior to a forward jump, the OCCO dihedral can perform $n_{bj}$ backward jumps in
the $t$ state. Thus, in general, sequences $g^\pm\!\rightarrow\!n_{bj}(t\!\rightarrow
\!g^\pm\rightarrow\!)\;t\!\rightarrow\! g^\mp$ lead to the exploration of all
conformational states. Provided the back-jump probability in the $t$ state is independent
of the history, the probability of finding a sequence with $n_{bj}$ backward jumps in the
$t$ state is given by $(1\!-\!p_b^{\,t})(p_b^{\,t})^{n_{bj}}$. Then, the average time
$\langle t_{gtg}\rangle$ to move from one of the $gauche$ states to the other can be
written as

\begin{eqnarray}\label{eq_t3}
\langle t_{gtg} \rangle &=&(1\!-\!p_b^{\,t})\sum_{n_{bj}=0}(p_b^{\,t})^{n_{bj}}(t_{fl}+n_{bj}\,t_{bf}) \nonumber \\
&=& t_{fl}+\frac{p_b^{\,t}}{1-p_b^{\,t}}\;t_{bf}\,.
\end{eqnarray}


Here, $t_{bf}$ is the time needed for a forward-backward jump sequence and $t_{fl}$ is
the sum of the times elapsing prior to the first and the last conformational transition.
When we assume that the waiting times can depend on the jump direction, but are otherwise
independent of the history, $t_{bf}$ and $t_{fl}$ are determined by the mean waiting
times $\langle t_w^{t,f}\rangle$ and $\langle t_w^{t,b}\rangle$ and by the mean waiting
time $\langle t_w^{g}\rangle$ in the $g^{\pm}$ states, which is found to be essentially
independent of the jump direction. Specifically, $t_{bf}\!=\!t_w^{t,b}\!+\!t_w^g$ and
$t_{fl}\!=\!t_w^g/2\!+\!t_w^{t,f}$. The factor 1/2 in the latter equation is a
consequence of the fact that, on average, the randomly chosen time origin lies in the
middle of the waiting time before the first jump.

Utilizing the knowledge of the mean waiting times and of the back-jump probabilities, we
calculate the times $\langle t_{gtg}\rangle$ according to Eq.\ (\ref{eq_t3}). In Fig.\
\ref{fig9}, we see that the temperature dependence of $\langle t_{gtg}\rangle$ deviates
from an Arrhenius law due to the higher back-jump probabilities at lower temperatures.
While the calculated and the actual exploration times are similar at high temperatures,
the former show a weaker temperature dependence. Therefore, we dropped the assumption
that direct transitions between the $gauche$ states can be neglected and recalculated the
mean times $\langle t_{gtg}\rangle$. In Fig.\ \ref{fig9}, it is evident that considering
the transitions between the $gauche$ states has hardly any effect. Therefore, we refrain
from specifying the equations for this case. Also, we expect that the assumption to start
in one of the $gauche$ states is not crucial, in particular at low temperatures, where
the occupation of the $trans$ state is small. The deviations between the calculated and
the actual exploration times rather show that the assumption of history independent
waiting times and back-jump probabilities is not justified, further illustrating the
complexity of the conformational dynamics.

\begin{figure}
\centering
\includegraphics[width=8.25cm]{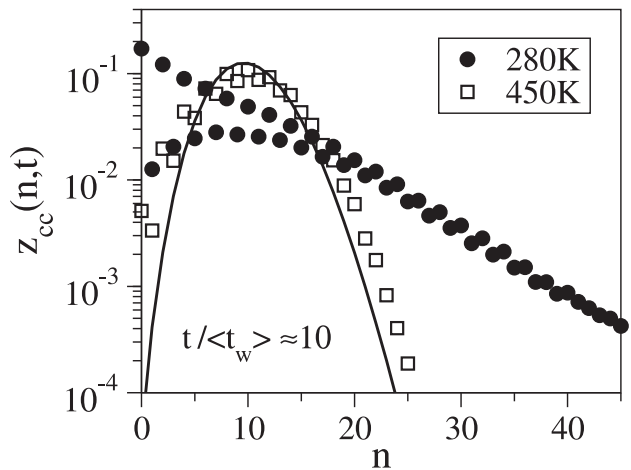}
\caption{\small \textsl{Probability $z_{cc}(n,t)$ of finding $n$ conformational
transitions of the OCCO dihedrals during a time interval $t$. We used
$t\!=\!125.5\mathrm{\,ps}$ and $t\!=\!39.5\mathrm{\,ps}$ for PEO at
$T\!=\!280\mathrm{\,K}$ ($\langle t_w\rangle\!=\!14.3\mathrm{\,ps}$) and
$T\!=\!450\mathrm{\,K}$ ($\langle t_w\rangle\!=\!3.9\mathrm{\,ps}$),
respectively. Thus, the ratio between the time interval and the mean waiting
time amounts to $t/t_w\!\approx\!10$ for both temperatures. The solid line is
the Poisson distribution calculated using the values of $t$ and $t_w$ at
$T\!=\!450\mathrm{\,K}$, i.e., $t/t_w\!=\!10.1$. }\normalsize}\label{fig13}
\end{figure}

Finally, we analyze the torsional motion of the OCCO dihedrals on a longer time scale
$t\!\approx\!10\langle t_w \rangle$, where $\langle t_w \rangle$ is the mean waiting time
resulting from all conformational transitions of this dihedral species. We calculate the
probability distribution $z_{cc}(n,t)$ of finding $n$ conformational transitions during a
time interval $t$. Figure \ref{fig12} shows the distributions $z_{cc}(n,t)$ obtained for
$t\!\approx\!10\langle t_w \rangle$ at $T\!=\!280\mathrm{\,K}$ and
$T\!=\!450\mathrm{\,K}$. The data for the higher temperature resemble the Poisson
distribution for the used ratio $t/\langle t_w \rangle\!\approx\!10$ and, hence, a Markov
process approximates the conformational dynamics. By contrast, substantial deviations
from the Poisson distribution are obvious for the lower temperature. Specifically,
$z_{cc}(n,t)$ is much broader, indicating that a large fraction of dihedrals performs
less or more transitions than expected for the case of uncorrelated jump events. Thus, at
sufficiently low temperatures, pronounced dynamical heterogeneities do not only govern
the structural relaxation, see Sec.\ \ref{sec_shd}, but also the conformational
relaxation.

Furthermore, it is evident from Fig.\ \ref{fig12} that it is more probable to find an
even than an uneven number of conformational transitions in a given time interval. This
effect, which is most pronounced for small $n$ and low temperatures, is another
consequence of the existence of correlated forward-backward jumps. It implies that, at a
given time, each OCCO dihedral has a preferred conformational state. After the exit of
its preferred state, a dihedral tends to return to this state within a very short period
of time performing a correlated backward jump. Then, on average, the dihedrals spend much
longer times in the preferred state than in the unpreferred state of the forward-backward
sequence, resulting in a higher probability of finding an even number of jumps,
consistent with the observation. At $T\!=\!280\mathrm{\,K}$, the difference between even
and uneven $n$ is small for $n\!>\!30$ and, hence, the information about the preferred
state vanishes after about 30 conformational transitions, i.e., after about 15
forward-backward jump sequences.

\subsection{Conclusions}

We have performed MD simulations to investigate the translational and rotational motion
of chemically realistic polymer models. For the studied PEO and PPO models, the
structural relaxation shows the typical properties of polymer melts, i.e., its time
dependence differs from a single exponential function and its temperature dependence
deviates from an Arrhenius law, in harmony with results in the literature.\cite{Bo_03_3}
Specifically, for both polymer models, a KWW function well interpolates the translational
and rotational correlation functions in the $\alpha$-relaxation regime, time temperature
superposition is obeyed, and a VFT law describes the temperature dependent
$\alpha$-relaxation times. The stretching and the temperature dependence of the
correlation functions are consistent with results from experimental work, confirming the
quality of the used force fields.

Recently, the applicability of MCT to the structural relaxation of chemically realistic
polymer models was controversially discussed.\cite{Kr_03,Pa_04,Co_07,Pa_06} Here, we have
performed a MCT analysis for the PPO model using the incoherent intermediate scattering
functions of the oxygen atoms. On the one hand, the analysis shows that a MCT power law
with a critical temperature $T_c\!=\!242\!\pm\!2\mathrm{\,K}$ well describes the
temperature dependent correlation times in the range $(T\!-\!T_c)/T_c\!=\!0.1\!-\!0.9$.
Also, for a momentum transfer $q$ corresponding to the intermolecular oxygen-oxygen
distance, further MCT predictions are fulfilled. On the other hand, the temperature
dependence of the $\alpha$-relaxation time deviates from a MCT power law in the immediate
vicinity of $T_c$ and the factorization theorem, being a central MCT prediction for the
$\beta$-relaxation regime, is violated. We conclude that MCT captures several aspects of
the structural relaxation at appropriate temperatures and length scales, however it does
not provide a complete description since energy barriers against the torsional motion
affect the dynamical behavior.

Furthermore, we have studied the heterogeneity and the cooperativity of the structural
relaxation. For the PEO and PPO models, we have demonstrated that highly mobile oxygen
atoms aggregate into transient clusters, indicating the spatially heterogeneous nature of
the dynamics. Furthermore, we have shown that cooperative string-like motion facilitates
the translational displacements of the highly mobile oxygen atoms at intermediate times
between ballistic motion and diffusive motion. Both clusters and strings increase in size
upon cooling and, hence, spatial heterogeneity and cooperativity are prominent aspects of
the molecular dynamics in particular at low temperatures, in harmony with results for
various models of glass-forming liquids.\cite{Gl_00_1} Concerning the transient nature of
these phenomena, we have found that the mean cluster size $S_w$ and the mean string
length $L_w$ are a maximum at comparable times $t_S\!\approx\!t_L$ in the
$\alpha$-relaxation regime for PEO and PPO. In this respect, the present findings differ
from previous results. Specifically, the clusters and the strings were reported to be
largest at significantly earlier times in the late-$\beta$/ early-$\alpha$ relaxation
regime for models of atomic liquids,\cite{Don_99_2,Ge_04} water\cite{Gi_03},
silica,\cite{Vo_04_1,Vo_04_2} and a bead-spring polymer.\cite{Ge_01,Ai_03} We conclude
that the existence of spatially heterogeneous and cooperative dynamics are common to a
broad variety of glass-forming liquids. However, the characteristics of these phenomena
differ among the materials and, hence, relating the properties of the heterogeneity and
the cooperativity of the molecular dynamics to the respective features of the structural
relaxation may yield interesting insights into the glass transition phenomenon.

For the PEO model, we have demonstrated that a straightforward study of the
conformational relaxation is possible, when mapping the continuous trajectories of the
dihedral angles onto discrete sequences of the dihedral states. Various results have
indicated a complex nature of the conformational dynamics. In particular, for both
dihedral species and for all dihedral states, the probability distributions of the
waiting times $t_w$, i.e., of the time intervals between two subsequent conformational
transitions, strongly deviate from an exponential function, indicating that a Markov
process does not apply to the conformational dynamics. To analyze the origin of this
behavior, we have discriminated between backward and forward jumps, i.e., we have
distinguished whether or not the dihedrals are in the same conformational state after
exactly two transitions. This analysis revealed that the nonexponential waiting-time
distributions are a consequence of correlated forward-backward jumps, which are an
important aspect of the conformational dynamics at sufficiently low temperatures.

Consistent with previous results for all-atom polymer models,\cite{Pa_04} we have
observed that the mean waiting times show a weaker temperature dependence than the time
constants of the conformational and structural relaxations. Hence, it is not possible to
explain the glassy slowdown on the basis of single events, but it is necessary to
correlate the times and the directions of several consecutive conformational transitions.
Analyzing the directions of subsequent transitions, we have shown that the probability of
backward jumps increases upon cooling and, hence, the exploration of the conformational
states is slower at lower temperatures due to both longer waiting times and higher
back-jump probabilities. To obtain insights into the times of consecutive transitions, we
considered probability distributions $z_x(n,t)$ of finding $n$ transitions in a time
interval $t$. In this way, we have revealed that the conformational dynamics resembles a
Poisson process at high temperatures. However, when the temperature is decreased,
dynamical heterogeneities become important for the conformational relaxation and the
dihedrals start having a preferred conformational state at a given time. After an exit of
the preferred state state, the dihedrals show a high tendency to return to this state via
a correlated backward jump. At the studied temperatures, the dihedrals remember this
preferred state for up to about 30 conformational transitions. In other words, the
conformational states are not sampled according to their statistical weights in the early
stages of the conformational relaxation.

\section{Acknowledgement}

The author thanks the Deutsche Forschungsgemeinschaft (DFG) for funding through grant VO
905/3-1 and A.\ Heuer for computer time.


\begin{thebibliography}{99}

\bibitem{Ed_96} Ediger, M.\ D.; Angell, C.\ A.; Nagel, S.\ R.\ \emph{J.\ Phys.\ Chem.\ }$\mathbf{1996}$, 100, 13200
\bibitem{Bi_00} Binder, K.\ \emph{J.\ Non-Cryst.\ Solids} $\mathbf{2000}$, 274, 332
\bibitem{Gl_00_1} Glotzer, S.\ C.\ \emph{J.\ Non-Cryst.\ Solids} $\mathbf{2000}$, 274, 342
\bibitem{De_01} Debenedetti, P.\ G., Stillinger, F.\ H.\ \emph{Nature}, $\mathbf{2001}$, 410, 259
\bibitem{Pa_04} Paul, W.; Smith, G.\ D.\ \emph{Rep.\ Prog.\ Phys.\ }$\mathbf{2004}$, 67, 1117
\bibitem{Ba_05} Baschnagel, J.; Varnik, F.\ \emph{J.\ Phys.: Condens.\ Matter} $\mathbf{2005}$, 17, R851
\bibitem{Sm_07} Smith, G.\ D.; Bedrov, D.\ \emph{J.\ Pol.\ Sci.: Part B: Pol.\ Phys.\ }$\mathbf{2007}$, 45, 627
\bibitem{Co_07} Colmenero, J.; Narros, A.; Alvarez, F.; Arbe, A.; Moreno, A.\ J.\ \emph{J.\ Phys.: Condens.\ Matter} $\mathbf{2007}$, 19, 205127
\bibitem{Go_92} G\"otze, W.; Sjogren, L.\ \emph{Rep.\ Prog.\ Phys.\ }$\mathbf{1992}$, 55, 241
\bibitem{Ko_95} Kob, W.; Andersen, H.\ C.\ \emph{Phys.\ Rev.\ E}  $\mathbf{1995}$, 52, 4134
\bibitem{Ho_01} Horbach, J.; Kob, W.\ \emph{Phys.\ Rev.\ E}  $\mathbf{2001}$, 64, 041503
\bibitem{Kr_03} Krushev, S.; Paul, W.; \emph{Phys.\ Rev.\ E} $\mathbf{2003}$, 67, 021806
\bibitem{Pa_06} Paul, W.; Bedrov, D.; Smith, G.\ D.\ \emph{Phys.\ Rev.\ E} $\mathbf{2006}$, 74, 021501
\bibitem{Ad_65} Adam, G.; Gibbs, J.\ H.\ \emph{J.\ Chem.\ Phys.\ }$\mathbf{1965}$, 43, 139
\bibitem{Ga_02} Garrahan, J.\ P.; Chandler, D.\ \emph{Phys.\ Rev.\ Lett.\ }$\mathbf{2002}$, 89, 035704
\bibitem{Bo_98} B\"ohmer, R.; Chamberlin, R.\ V.; Diezemann, G.; Geil, B.; Heuer, A.; Hinze, G.; Kuebler, S.\ C.; Richert, R.; Schiener, B.; Sillescu, H.; Spiess, H.\ W.; Tracht, U.; Wilhelm, M; \emph{J.\ Non-Cryst.\ Solids} $\mathbf{1998}$, 235, 1
\bibitem{Si_99} Sillescu, H.\ \emph{J.\ Non-Cryst.\ Solids} $\mathbf{1999}$, 243, 81
\bibitem{Ed_00} Ediger, M.\ D.\ \emph{Annu.\ Rev.\ Phys.\ Chem.\ }$\mathbf{2000}$, 51, 99
\bibitem{Tr_98} Tracht, U.; Wilhelm, M.; Heuer, A.; Feng, H.; Schmidt-Rohr, K.; Spiess, H.\ W.\ \emph{Phys.\ Rev.\ Lett.\ }$\mathbf{1998}$, 81, 2727
\bibitem{Re_01} Reinsberg, S.\ A.; Qiu, X.\ H.; Wilhelm, M.; Spiess, H.\ W.; Ediger M.\ D.\ \emph{J.\ Chem.\ Phys.\ }$\mathbf{2001}$, 114, 7299
\bibitem{Qi_03} Qiu, X.\ H.; Ediger M.\ D.\ \emph{J.\ Phys.\ Chem.\ B} $\mathbf{2003}$, 107, 459

\bibitem{Ko_97} Kob, W.; Donati, C.; Plimpton, S.\ J.; Poole, P.\ H.; Glotzer, S.\ C.\ \emph{Phys.\ Rev.\ Lett.\ }$\mathbf{1997}$, 79, 2827
\bibitem{Dol_98} Doliwa, B.; Heuer, A.\ \emph{Phys.\ Rev.\ Lett.\ }$\mathbf{1998}$, 80, 4915
\bibitem{Don_99_1} Donati, C.; Glotzer, S.\ C.; Poole, P.\ H.\ \emph{Phys.\ Rev.\ Lett.\ }$\mathbf{1999}$, 82, 5064
\bibitem{Don_99_2} Donati, C.; Glotzer, S.\ C.; Poole, P.\ H.; Kob, W.; Plimpton, S.\ J.\ \emph{Phys.\ Rev.\ E} $\mathbf{1999}$, 60, 3107
\bibitem{Be_99} Bennemann, C.; Donati, C.; Baschnagel, J.; Glotzer, S.\ C.\ \emph{Nature} $\mathbf{1999}$, 399, 246
\bibitem{Gl_00_2} Glotzer, S.\ C.; Novikov, V.\ N.; Schr{\o}der, T.\ B.\ \emph{J.\ Chem.\ Phys.\ }$\mathbf{2000}$, 112, 509
\bibitem{Ge_01} Gebremichael, Y.; Schr{\o}der, T.\ B.; Starr, F.\ W.; Glotzer, S.\ C.\ \emph{Phys.\ Rev.\ E} $\mathbf{2001}$, 64, 051503
\bibitem{Ge_04} Gebremichael, Y.; Vogel, M.; Glotzer, S.\ C.\ \emph{J.\ Chem.\ Phys.\ }$\mathbf{2004}$, 120, 4415
\bibitem{Don_98} Donati, C.; Douglas, J.\ F.; Kob, W.; Plimpton, S.\ J.; Poole, P.\ H.; Glotzer, S.\ C.\ \emph{Phys.\ Rev.\ Lett.\ }$\mathbf{1998}$, 80, 2338
\bibitem{Ai_03} Aichele, M.; Gebermichael, Y.; Starr, F.\ W.; Baschnagel, J.; Glotzer S.\ C.\ \emph{J.\ Chem.\ Phys.\ }$\mathbf{2003}$, 119, 5290
\bibitem{Qi_99} Qian, J.; Hentschke, R.; Heuer, A.\ \emph{J.\ Chem.\ Phys.\ }$\mathbf{1999}$, 110, 4514
\bibitem{Gi_03} Giovambattista, N.; Buldyrev, S.\ V.; Starr, F.\ W.; Stanley, H.\ E.\ \emph{Phys.\ Rev.\ Lett.\ }$\mathbf{2003}$, 90, 085506
\bibitem{Vo_04_1} Vogel, M.; Glotzer, S.\ C.\ \emph{Phys.\ Rev.\ Lett.\ }$\mathbf{2004}$, 92, 255901
\bibitem{Vo_04_2} Vogel, M.; Glotzer, S.\ C.\ \emph{Phys.\ Rev.\ E} $\mathbf{2004}$, 70, 061504
\bibitem{Te_04} Teboul, V.; Monteil, A.; Fai, L.\ C.; Kerrache, A.; Maabou, S.\ \emph{Eur.\ Phys.\ J.\ B} $\mathbf{2004}$, 40, 49

\bibitem{Bo_00} Borodin, O.; Smith, G.\ D.\ \emph{Macromolecules} $\mathbf{2000}$, 33, 2273
\bibitem{Bo_03_0} Borodin, O.; Smith, G.\ D.; Bandyopadhyaya, R.; Byutner, O.\ \emph{Macromolecules} $\mathbf{2003}$, 36, 7873
\bibitem{Gr_91} Gray, F.\ M.\ \ \emph{Solid Polymer Electrolytes, Wiley, New York} $\mathbf{1991}$
\bibitem{Ne_95} Neyertz, S.; Brown, D.\ \emph{J.\ Chem.\ Phys.\ }$\mathbf{1995}$, 102, 9725
\bibitem{Mu_95} M\"uller-Plathe, F.; van Gunsteren, W.\ F.\ \emph{J.\ Chem.\ Phys.\ }$\mathbf{1995}$, 103, 4745
\bibitem{Ca_95} Catlow, C.\ R.\ A.; Mills, G.\ E.\ \emph{Electrochim.\ Acta} $\mathbf{1995}$, 40, 2057
\bibitem{Li_96} Lin, B.; Boinske, P.\ T.; Halley, J.\ W.\ \emph{J.\ Chem.\ Phys.\ }$\mathbf{1996}$, 105, 1668
\bibitem{Sm_96} Smith, G.\ D.; Yoon, D.\ Y.; Jaffe, R.\ L.; Colby, R.\ H.; Krishnamoorti, R.; Fetters, L.\ J.\ \emph{Macromolecules} $\mathbf{1996}$, 29, 3462
\bibitem{Ah_00} Ahlstr\"om, P.; Borodin, O.; Wahnstr\"om, G.; Wensink, E.\ J.\ W.; Carlsson, P.; Smith, G.\ D.\ \emph{J.\ Chem.\ Phys.\ }$\mathbf{2000}$, 112, 10669
\bibitem{Bo_03_2} Borodin, O.; Douglas, R.; Smith, G.\ D.; Trouw, F.; Petrucci, S.\ \emph{J.\ Phys.\ Chem.\ B} $\mathbf{2003}$, 107, 6813
\bibitem{Bo_03_3} Borodin, O.; Smith, G.\ D.; Douglas, R.\ \emph{J.\ Phys.\ Chem.\ B} $\mathbf{2003}$, 107, 6824
\bibitem{Ha_00} Hackett, E.; Manias, E.; Giannelis, E.\ P.\ \emph{Chem.\ Mater.\ }$\mathbf{2000}$, 12, 2161
\bibitem{Ku_03} Kuppa, V.; Manias, E.\ \emph{J.\ Chem.\ Phys.\ }$\mathbf{2003}$, 118, 3421

\bibitem{Bo_03_1} Borodin, O.; Smith, G.\ D.\ \emph{J.\ Phys.\ Chem.\ B} $\mathbf{2003}$, 107, 6801
\bibitem{Sm_98} Smith, G.\ D.; Borodin, O.; Bedrov, D.\  \emph{J.\ Phys.\ Chem.\ A} $\mathbf{1998}$, 102, 10318
\bibitem{FN} In Ref.\ \onlinecite{Bo_03_1}, the correct values of the torsional parameters $k_3$ for the OCCH and HCCH dihedrals are -0.28 each.
\bibitem{GROM} Lindahl E.; Hess, B.; van der Spoel D.\ \emph{J.\ Mol.\ Mod.\ }$\mathbf{2001}$, 7, 306; Berendsen, H.\ J.\ C.; van der Spoel D.; van Drunen, R.\ \emph{Comp.\ Phys.\ Comm.\ }$\mathbf{1995}$, 91, 43
\bibitem{LINCS} Hess, B.; Bekker, H.; Berendsen, H.\ J.\ C.; Fraaije, J.\ G.\ E.\ M.\ \emph{J.\ Comp.\ Chem.\ }$\mathbf{1997}$, 18, 1463
\bibitem{PME} Essman, U.; Perela, L.; Berkowitz, M.\ L.; Darden, T.; Lee H.; Pedersen, L.\ G.\ \emph{J.\ Chem.\ Phys.\ }$\mathbf{1995}$, 103, 8577
\bibitem{RP} Parrinello, M.; Rahman, A.\ \emph{J.\ Appl.\ Phys.\ }$\mathbf{1981}$, 52, 7182
\bibitem{NH} Nos\'{e}, S.\ \emph{Mol.\ Phys.\ }$\mathbf{1984}$, 52, 255; Hoover, W.\ G.\ \emph{Phys.\ Rev.\ A} $\mathbf{1985}$, 31, 1695
\bibitem{Vo_06} Vogel, M.; Torbr\"ugge, T.\ \emph{J.\ Chem.\ Phys.\ }$\mathbf{2006}$, 125, 164901
\bibitem{VFT} Vogel, H.\ \emph{Z.\ Phys.\ }$\mathbf{1921}$, 22, 645; Fulcher, G.\ S.\ \emph{J.\ Am.\ Ceram.\ Soc.\ }$\mathbf{1925}$, 8, 339; Tammann, G.; Hesse, W., \emph{Z.\ Anorg.\ Allg.\ Chem.\ }$\mathbf{1926}$, 156, 245
\bibitem{Be_97} Bergman, R.; B\"orjesson, L.; Torell, L.\ M.; Fontana, A.\ \emph{Phys.\ Rev.\ B} $\mathbf{1997}$, 56, 11619
\bibitem{Le_99} Leon, C.; Ngai, K.\ L.; Roland, C.\ M.\ \emph{J.\ Chem.\ Phys.\ }$\mathbf{1999}$, 110, 11585
\bibitem{Si_91} Sidebottom, D.\ L.; Johari G.\ P.\ \emph{J.\ Pol.\ Sci.: Part B} $\mathbf{1991}$, 29, 1215
\bibitem{Gl_00} Gleim, T.; Kob, W.\ \emph{Eur.\ Phys.\ J.\ B} $\mathbf{2000}$, 13, 83
\bibitem{Si_92} Sidebottom, D.\ L.; Bergman, R.; B\"orjesson, L.; Torell, L.\ M.\ \emph{Phys.\ Rev.\ Lett.\ }$\mathbf{1992}$, 68, 3587
\bibitem{Sm_97} Smith, G.\ D.; Yoon, D.\ Y.; Wade, C.\ G.; O'Leary D.; Chen, A.; Jaffe,  R.\ L.\ \emph{J.\ Chem.\ Phys.\ }$\mathbf{1997}$, 106, 3798

\end{thebibliography}
\end{document}